\documentclass[aps,showpacs,a4paper,floatfix]{revtex4}
\usepackage[dvips]{graphicx}

\newcommand{\balpha}{{\mbox{\boldmath$\alpha$}}}

\newcommand{\be}{\begin{eqnarray}}
\newcommand{\ee}{\end{eqnarray}}
\newcommand{\la}{\langle}
\newcommand{\ra}{\rangle}
\newcommand{\bfx}{{\bf x}}
\newcommand{\bfy}{{\bf y}}

\newcommand{\bfk}{{\bf k}}

\newcommand{\veps}{\varepsilon}
\newcommand{\eps}{\epsilon}
\newcommand{\beps}{{\mbox{\boldmath$\epsilon$}}}
\begin{document}
\title{Interelectronic-interaction effect on
the transition probability in high-$Z$ He-like ions}
\author{ P. Indelicato$^1$, V. M. Shabaev$^2$,
and A. V. Volotka$^2$ }

\affiliation{
$^1$Laboratoire Kastler-Brossel, 
\'Ecole Normale Sup\' erieure et Universit\' e P. et M. Curie, \\
 Case 74, 4, place Jussieu,
75252 Paris CEDEX 05, France\\
$^2$Department of Physics, St.Petersburg State University,
Oulianovskaya 1, Petrodvorets, St.Petersburg 198504, Russia\\
}
\begin{abstract}

The interelectronic-interaction effect on the transition probabilities in
high-$Z$ He-like ions is  investigated within a systematic quantum
electrodynamic approach. The calculation formulas for the  
interelectronic-interaction corrections
 of first order in $1/Z$ are derived 
using the two-time Green function method. These formulas are
employed for 
numerical evaluations of the magnetic transition probabilities 
in heliumlike ions. The results of the calculations 
are compared with experimental values and previous calculations.

\end{abstract}

%32.70.Cs, 31.25. -v, 31.30.Jv
\maketitle
%\newpage

\section{Introduction}

During the last few years, transition probabilities in 
 heliumlike ions were calculated by
a number of authors \cite{lin90,joh95,ind96,der98}.
In these calculations,
 the interelectronic-interaction
effects on the transition probabilities were accounted for by
employing the relativistic many-body perturbation theory (RMBPT) 
\cite{lin90,joh95,der98}  and 
the  multiconfiguration Dirac-Fock (MCDF) method \cite{ind96}.
Since these methods are based on using the Coulomb-Breit hamiltonian,
they have to deal with 
a separate treatment of the positive- and 
negative-energy state contributions.
As was first indicated in \cite{ind96}, 
the contribution from the negative-continuum contribution is 
very sensitive to 
the choice of the one-electron model potential, which
is used as the starting point of any RMBPT or MCDF calculation.
In particular, using a standard Dirac-Fock approximation, in 
\cite{ind96} it has been demonstrated
 that to achieve the agreement between theory and
experiment for the magnetic dipole transition
$2^3S_1 \rightarrow 1^1S_0$ it is necessary to take into account
both correlation and negative-continuum effects. 
This statement is closely related to a problem of significant
numerical cancellations that may occur in low-Z systems, if
an improper one-electron approximation is used. For a rigorous QED
approach for low-Z systems and corresponding calculations
 we refer to \cite{lac01,pac03}. 

The main goal of the present paper is to perform a complete QED
calculation of the 
interelectronic-interaction correction of first order in $1/Z$ to
the magnetic transition probabilities in high-$Z$ He-like ions.
To derive the calculation formulas for these corrections from
the first principles of QED we use the two-time Green function
method developed in \cite{shab90a,shab90b,shab94} and described
in details in \cite{shab02}. In Sec. \ref{sec:base}, we formulate the basic
equations of this method for the case of nondegenerate states
and apply it for the derivation of the desired formulas.
The numerical results for the transitions $2^3S_1 \rightarrow 1^1S_0$,
$\ 2^3P_2 \rightarrow 1^1S_0$, and $3^3S_1 \rightarrow 2^3S_1$
are presented in Sec. \ref{sec:res}. Both Feynman and Coulomb gauges are
used for the photon propagator to demonstrate
the gauge independence of the final results. 
The results of the calculations are compared with
previous theoretical results and  with experiment.

The relativistic units ($\hbar=c=1$) and
the Heaviside charge unit ( $\alpha=e^2/(4\pi),e<0$ )
are used in the paper.

\section{Basic formulas}
\label{sec:base}
We consider the transition of a high-$Z$ two-electron ion
from an initial state $a$ to a final state $b$ with the emission
of a photon with momentum $\bfk_f$ and polarization $\beps_f$.
The transition probability is given as 
\be
dW=2\pi |\tau_{\gamma_f,b;a}|^2\delta(E_b+k_f^0-E_a)d\bfk_f\,,
\ee
where $\tau_{\gamma_f,b;a}$ is the transition amplitude
which is connected with the $S$-matrix element by
\be \label{deftau}
\la \bfk_f, \beps_f;b|S|a\ra=2\pi i \tau_{\gamma_f,b;a}
\delta(\veps_b+k_f^0-\veps_a)\,, 
\ee
and $k_{f}^{0} \equiv |{\bf k}_{f}|$.

We assume that in zeroth (one-electron) approximation
the initial and final states of the ion are described
by one-determinant wave functions
\be \label{ua}
u_a(\bfx_1,\bfx_2)=\frac{1}{\sqrt{2}}\sum_{P}(-1)^P
\psi_{Pa_1}(\bfx_1)
\psi_{Pa_2}(\bfx_2)\,, \\
u_b(\bfx_1,\bfx_2)=\frac{1}{\sqrt{2}}\sum_{P}(-1)^P
\psi_{Pb_1}(\bfx_1)
\psi_{Pb_2}(\bfx_2)\,.
\label{ub}
\ee
To describe the process under consideration we introduce
the Green function
$g_{\gamma_f,b;a}(E',E)$ by
\begin{eqnarray}  \label{transgsmall}     
 g_{\gamma_f,b;a}(E',E) \delta(E'+k^0-E)  &=&
 \frac{1}{2!}
     \int_{-\infty}^{\infty}dp_{1}^{0}dp_{2}^{0}
    dp_{1}^{\prime 0}dp_{2}^{\prime 0}\nonumber\\
&& \times \delta(E-p_{1}^{0}  - p_{2}^{0})
            \delta(E'-p_{1}^{\prime 0} - p_{2}^{\prime 0}) 
\nonumber\\  
 && \times \int d\bfx_1 d\bfx_2 d\bfx_1' d\bfx_2'
 u^{\dag}_{b}(\bfx_1',\bfx_2')\nonumber\\
&&\times
  G_{\gamma_f}
((p_{1}^{\prime 0},\bfx_1'), (p_{2}^{\prime 0},\bfx_2')
;k^0;(p_{1}^{0},\bfx_1)
 ,(p_{2}^{0},\bfx_2))
\gamma_{1}^{0} \gamma_{2}^{0}
 u_{a}(\bfx_1,\bfx_2)      
\,,
       \label{gtrans22}
\end{eqnarray}
where
\be \label{gtrans23}
\lefteqn{  G_{\gamma_f}
((p_{1}^{\prime 0},\bfx_1') ,(p_{2}^{\prime 0},\bfx_2');k^0;
(p_{1}^{0},\bfx_1),(p_{2}^{0},\bfx_2))} \nonumber\\
&=&\frac{2\pi}{i}\frac{1}{(2\pi)^{5}}
\int_{-\infty}^{\infty}dx_{1}^{0}dx_{2}^{0}
dx_{1}^{\prime 0}dx_{2}^{\prime 0} \int d^4y\nonumber\\
&&\times\exp{(ip_{1}^{\prime 0}x_{1}^{\prime 0}+ip_{2}^{\prime 0}x_{2}^
{\prime 0}-ip_{1}^{ 0}x_{1}^{0}-ip_{2}^{ 0}
x_{2}^{0}+ik^0y^0)}\nonumber\\
&&\times A_f^{\nu *}(\bfy)
\la 0|T\psi(x_{1}^{\prime}) \psi(x_{2}^{\prime})j_{\nu}(y) 
\overline {\psi}(x_{2})\overline {\psi} (x_{1})|0\ra\, 
\ee
is the Fourier transform of the four-time Green function
describing the process,
$\psi(\bfx)$ is the electron-positron field operator in
the Heisenberg representation, and
\begin{eqnarray}\label{defA}
A_{f}^{\nu}(\bfx)=
 \frac{{\eps}_{f}^{\nu} \exp{(i{\bf k}_{f}\cdot{\bf x})}}
{\sqrt{2k_{f}^{0}(2\pi )^3}}\,
\end{eqnarray}
is the wave function of the emitted photon.
The transition amplitude 
$S_{\gamma_f,b;a}\equiv \la \bfk_f, \beps_f;b|S|a\ra$ 
is calculated by \cite{shab90a,shab90b,shab02}
\be \label{transfin}
S_{\gamma_f,b;a}
&=&Z_3^{-1/2}\delta(E_b+k_f^0-E_a)
\oint_{\Gamma_b}dE'\oint_{\Gamma_a}dE g_{\gamma_f,b;a}
(E',E)\nonumber\\
&&\times \Bigl[\frac{1}{2\pi i}\oint_{\Gamma_b}dE g_{bb}(E)\Bigr]^{-1/2}
\Bigl[\frac{1}{2\pi i}\oint_{\Gamma_a}dE g_{aa}(E)\Bigr]^{-1/2}
\,.
\ee
Here $g_{aa}(E)$ is defined by
\begin{eqnarray}       
 g_{aa}(E) \delta(E'-E)  &=&
 \frac{2\pi}{i}\frac{1}{2!}
     \int_{-\infty}^{\infty}dp_{1}^{0}dp_{2}^{0}
    dp_{1}^{\prime 0}dp_{2}^{\prime 0}\nonumber\\
&& \times \delta(E-p_{1}^{0}  - p_{2}^{0})
            \delta(E'-p_{1}^{\prime 0} - p_{2}^{\prime 0}) 
\nonumber\\  
 && \times \int d\bfx_1 d\bfx_2 d\bfx_1' d\bfx_2'
%  \nonumber\\ && \quad 
  \times u^{\dag}_{a}(\bfx_1',\bfx_2')
  G\left((p_{1}^{\prime 0},\bfx_1'), (p_{2}^{\prime 0},\bfx_2');
(p_{1}^{0},\bfx_1)
 ,(p_{2}^{0},\bfx_2)\right)
\gamma_{1}^{0} \gamma_{2}^{0}
 u_{a}(\bfx_1,\bfx_2)      
\,,
       \label{gtrans22a}
\end{eqnarray}
where
\be \label{gtrans23a}
\lefteqn{  G
((p_{1}^{\prime 0},\bfx_1') ,(p_{2}^{\prime 0},\bfx_2');
(p_{1}^{0},\bfx_1),(p_{2}^{0},\bfx_2))} \nonumber\\
&=&\frac{1}{(2\pi)^{4}}
\int_{-\infty}^{\infty}dx_{1}^{0}dx_{2}^{0}
dx_{1}^{\prime 0}dx_{2}^{\prime 0}
\;\exp{(ip_{1}^{\prime 0}x_{1}^{\prime 0}+ip_{2}^{\prime 0}x_{2}^
{\prime 0}-ip_{1}^{ 0}x_{1}^{0}-ip_{2}^{ 0}
x_{2}^{0})}\nonumber\\
&&
\times \la 0|T\psi(x_{1}^{\prime}) \psi(x_{2}^{\prime})
\overline {\psi}(x_{2})\overline {\psi} (x_{1})|0\ra\,
\ee 
is the Fourier transform of the four-time Green function
describing the ion; $g_{bb}(E)$ is defined by a similar
equation.
The contours $\Gamma_a$ and $\Gamma_b$ surround the
poles corresponding to the initial and final levels and
keep outside all other singularities of the Green  functions.
It is assumed that they are oriented anticlockwise.
The  Green functions $G$ and $G_{\gamma_f}$ are
constructed by perturbation theory after the transition to
the interaction representation and using Wick's theorem.
The Feynman rules for $G$ and  $G_{\gamma_f}$ are given,
e.g., in \cite{shab02}.

Below we consider the transition probability in high-Z He-like ion
to zeroth and first order in $1/Z$.

\subsection{Zeroth order approximation}

\begin{figure}
\includegraphics{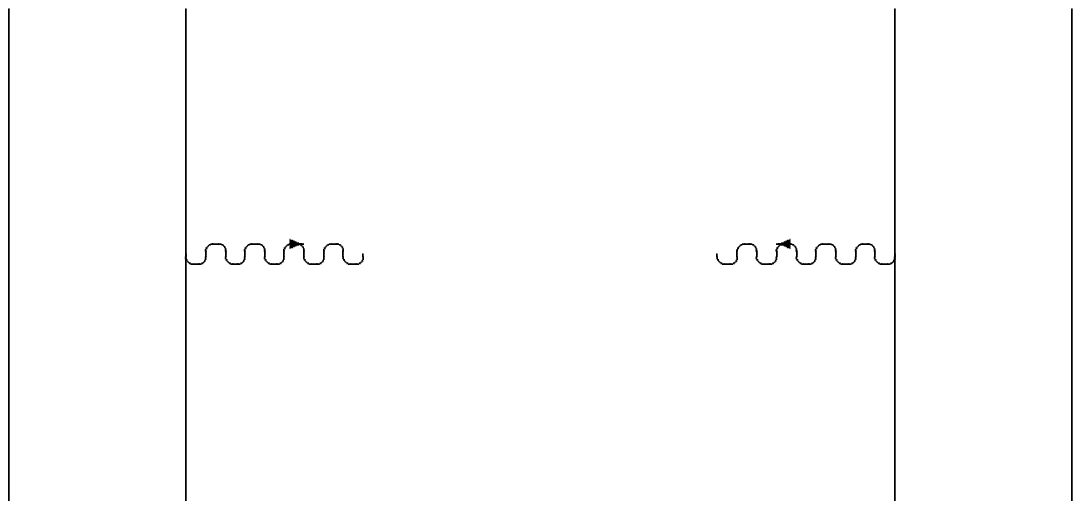}
\caption{The photon emission  by a
heliumlike ion in zeroth order approximation.\label{fig:zero}}
\end{figure}

To  zeroth order in $1/Z$ the transition amplitude
is described by the diagrams shown in Fig. 1\ref{fig:zero}.
Formula (\ref{transfin}) gives 
\begin{eqnarray}\label{zer1}
S^{(0)}_{\gamma_f,b;a}=\delta(E_{b}
+k_{f}^0-E_{a})\oint_{\Gamma_b}dE' \oint_{\Gamma_a}dE 
\, g^{(0)}_{\gamma_{f},b;a}(E',E)\,,
\end{eqnarray}
where the superscript indicates the order in $1/Z$.
According to  definition (\ref{transgsmall})
and the Feynman rules for $G_{\gamma_f}$ \cite{shab02},
we have 
\be
\lefteqn{g^{(0)}_{\gamma_f,b;a}(E',E)
\delta(E'+k^0-E)}\nonumber\\
&&=\sum_{P}(-1)^P \int_{-\infty}^{\infty}
dp_1^0 dp_2^0 dp_1'^0 dp_2'^0
\delta(E-p_1^0-p_2^0)
\delta(E'-p_1'^0-p_2'^0)
\nonumber\\
&&\times \Bigl\{\la Pb_1|\frac{i}{2\pi}
\sum_{n_1}\frac{|n_1\ra\la n_1|}{p_1'^0-\veps_{n_1}
(1-i0)}\frac{2\pi}{i}e\alpha_{\mu}\delta(p_1'^0+
k^0-p_1^0) A_f^{\mu *}\nonumber\\
&&\times\frac{i}{2\pi}
\sum_{n_2}\frac{|n_2\ra\la n_2|}{p_1^0-\veps_{n_2}
(1-i0)}|a_1\ra
\la Pb_2|\frac{i}{2\pi}
\sum_{n_3}\frac{|n_3\ra\la n_3|}{p_2^0-\veps_{n_3}
(1-i0)}|a_2\ra \delta(p_2'^0-p_2^0)\nonumber\\
&&+ 
\la Pb_1|\frac{i}{2\pi}
\sum_{n_1}\frac{|n_1\ra\la n_1|}{p_1^0-\veps_{n_1}
(1-i0)}|a_1\ra \delta(p_1'^0-p_1^0)\nonumber\\
&&\times
\la Pb_2|\frac{i}{2\pi}
\sum_{n_2}\frac{|n_2\ra\la n_2|}{p_2'^0-\veps_{n_2}
(1-i0)}\frac{2\pi}{i}e\alpha_{\mu}\delta(p_2'^0+
k^0-p_2^0) A_f^{\mu *}
\frac{i}{2\pi}
\sum_{n_3}\frac{|n_3\ra\la n_3|}{p_2^0-\veps_{n_3}
(1-i0)}|a_2\ra
\Bigr\}\,,
\ee
where $\alpha^{\mu}=\gamma^0 \gamma^{\mu}=(1,\balpha)$.
One obtains
\be
g^{(0)}_{\gamma_f,b;a}(E',E)&=&
\Bigl(\frac{i}{2\pi}\Bigr)^2
\int_{-\infty}^{\infty}dp_1^0
\frac{1}{p_1^0-(E-E')-\veps_{b_1}+i0}
\la b_1|e\alpha_{\mu}A_f^{\mu *}|a_1\ra
\frac{1}{p_1^0-\veps_{a_1}+i0}\frac{\delta_{a_2 b_2}}
{E-p_1^0-\veps_{a_2}+i0}
\nonumber\\
&&+\Bigl(\frac{i}{2\pi}\Bigr)^2
\int_{-\infty}^{\infty}dp_2^0
\frac{1}{p_2^0-(E-E')-\veps_{b_2}+i0}
\la b_2|e\alpha_{\mu}A_f^{\mu *}|a_2\ra
\frac{1}{p_2^0-\veps_{a_2}+i0}\frac{\delta_{a_1 b_1}}
{E-p_2^0-\veps_{a_1}+i0}\\
&=&\frac{i}{2\pi}
\frac{1}{E'-E_{b}^{(0)}}
[\la b_1|e\alpha_{\mu}A_f^{\mu *}|a_1\ra\delta_{a_2 b_2}
+\la b_2|e\alpha_{\mu}A_f^{\mu *}|a_2\ra\delta_{a_1 b_1}]
\frac{1}{E-E_a^{(0)}}
\,,
\ee
where 
$E_a^{(0)}=\veps_{a_1}+\veps_{a_2}$ and
$E_b^{(0)}=\veps_{b_1}+\veps_{b_2}$.
Substituting this expression into equation (\ref{zer1})
and integrating over $E$ and $E'$  we find
\be
S_{\gamma_f,b;a}^{(0)}=-2\pi i \delta(E_b+k_f^0-E_a)
[\la b_1|e\alpha_{\mu}A_f^{\mu *}|a_1\ra\delta_{a_2 b_2} +
\la b_2|e\alpha_{\mu}A_f^{\mu *}|a_2\ra\delta_{a_1 b_1}]
\ee
or, according to  definition (\ref{deftau}),
\be
\tau_{\gamma_f,b;a}^{(0)}=-\bigl[
\la b_1|e\alpha_{\mu}A_f^{\mu *}|a_1\ra\delta_{a_2 b_2} +
\la b_2|e\alpha_{\mu}A_f^{\mu *}|a_2\ra\delta_{a_1 b_1}
\bigr]\,.
\ee
The corresponding transition probability is
\begin{eqnarray}\label{inf10ab}
dW^{(0)}=
2\pi |\tau_{\gamma_f,b;a}^{(0)}|^2 \delta(E_b+k_f^0-E_a)
d\bfk_f\,.
\end{eqnarray}

\subsection{Interelectronic-interaction corrections
of first order in 1/Z}

The interelectronic-interaction corrections to the  transition
amplitude of first order 
in $1/Z$ are defined by diagrams shown in Fig. 2a,b.
Formula (\ref{transfin}) yields in the order under consideration
\begin{eqnarray}\label{transfirst1}
S^{(1)}_{\gamma_f,b;a}&=&\delta(E_{b}
+k_{f}^0-E_a)\;\Bigl[
\oint_{\Gamma_b}dE' \oint_{\Gamma_a}dE
 \,g_{\gamma_{f},b;a}^{(1)}(E',E)\nonumber\\
&&-\frac{1}{2}
\oint_{\Gamma_b}dE'\oint_{\Gamma_a}dE
 \, g_{\gamma_{f},b;a}^{(0)}(E',E)\;
\Bigl(
\frac{1}{2\pi i}\oint_{\Gamma_a} dE\, g_{aa}^{(1)}(E)
+\frac{1}{2\pi i}\oint_{\Gamma_b} dE\, g_{bb}^{(1)}(E)\Bigr)
 \Bigr]\,,
\end{eqnarray}
where $g_{aa}^{(1)}(E)$ and 
$g_{bb}^{(1)}(E)$ 
are defined by the first order
interelectronic-interaction diagram (Fig. 3).
Let us consider first the contribution of the diagrams
shown in Fig. 2a.
According to the definition (\ref{transgsmall})
and the Feynman rules for $G_{\gamma_f}$ \cite{shab02}, we have 
\be
\lefteqn{g^{(1a)}_{\gamma_f,b;a}(E',E)\delta(E'+k^0-E)
\;\;\;\;\;\;\;\;\;\;\;\;\;\;\;\;\;\;\;\;\;\;\;\;\;\;\;}
\nonumber\\
&&=\sum_{P}(-1)^P \int_{-\infty}^{\infty}
dp_1^0 dp_2^0 dp_1'^0 dp_2'^0
\delta(E-p_1^0-p_2^0)
\delta(E'-p_1'^0-p_2'^0)
\nonumber\\
&&\times \Bigl(\frac{i}{2\pi}\Bigr)^3
\int_{-\infty}^{\infty} dq^0 d\omega
\Bigl\{
 \frac{1}{p'^0_1-\veps_{Pb_1}+i0}
 \sum_n \la Pb_1 |e \alpha_{\mu} A_f^{\mu *}|n\ra
 \frac{1}{q^0-\veps_{n}(1-i0)}\nonumber\\
&&\times \la n Pb_2|I(\omega)|a_1 a_2\ra
 \frac{1}{p^0_1-\veps_{a_1}+i0}
 \frac{1}{p'^0_2-\veps_{Pb_2}+i0}
 \frac{1}{p^0_2-\veps_{a_2}+i0}\nonumber\\
&&\times \delta(p^0_1-\omega-q^0)
\delta(q^0-k^0-p_1'^0)
\delta(p^0_2+\omega-p_2'^0)\nonumber\\
&&+ \frac{1}{p'^0_2-\veps_{Pb_2}+i0}
 \sum_n \la Pb_2 |e \alpha_{\mu} A_f^{\mu *}|n\ra
 \frac{1}{q^0-\veps_{n}(1-i0)}\nonumber\\
&&\times \la Pb_1 n|I(\omega)|a_1 a_2\ra
 \frac{1}{p^0_2-\veps_{a_2}+i0}
 \frac{1}{p'^0_1-\veps_{Pb_1}+i0}
 \frac{1}{p^0_1-\veps_{a_1}+i0}\nonumber\\
&&\times \delta(p^0_2-\omega-q^0)
\delta(q^0-k^0-p_2'^0)
\delta(p^0_1+\omega-p_1'^0)\Bigr\}\,,
\ee
where $I(\omega)=e^2\alpha^{\mu}\alpha^{\nu}D_{\mu \nu}
(\omega)$ and
 \begin{eqnarray}
            D_{\rho\sigma} (\omega,{\bf x}-{\bf y})=
              -g_{\rho\sigma}\int \;\frac{d{\bf k}}
                   {(2\pi)^{3}}\;
\frac{\exp{(i{\bf k}\cdot({\bf x}-{\bf y}))}}
                           {\omega^{2}-{\bf k}^{2}+i0}             
         \label{e2e7}
     \end{eqnarray} 
is the photon propagator in the Feynman gauge.
One finds
\be \label{transint2}
g^{(1a)}_{\gamma_f,b;a}(E',E)&=&
\Bigl(\frac{i}{2\pi}\Bigr)^3
\sum_{P}(-1)^P \sum_n
\int_{-\infty}^{\infty} dp^0_2 dp_2'^0
\frac{1}{p'^0_2-\veps_{Pb_2}+i0}
\frac{1}{E'-p'^0_2-\veps_{Pb_1}+i0}\nonumber\\
&&\times
\frac{1}{p^0_2-\veps_{a_2}+i0}
\frac{1}{E-p^0_2-\veps_{a_1}+i0}
\la Pb_1|e \alpha_{\mu} A_f^{\mu *}|n\ra\nonumber\\
&&\times
\frac{1}{E-p_2'^0-\veps_n(1-i0)}
\la n Pb_2|I(p'^0_2-p_2^0)|a_1 a_2\ra
\nonumber\\
&&+\Bigl(\frac{i}{2\pi}\Bigr)^3
\sum_{P}(-1)^P \sum_n
\int_{-\infty}^{\infty} dp^0_1 dp_1'^0
\frac{1}{p'^0_1-\veps_{Pb_1}+i0}
\frac{1}{E'-p'^0_1-\veps_{Pb_2}+i0}\nonumber\\
&&\times
\frac{1}{p^0_1-\veps_{a_1}+i0}
\frac{1}{E-p^0_1-\veps_{a_2}+i0}
\la Pb_2|e \alpha_{\mu} A_f^{\mu *}|n\ra\nonumber\\
&&\times
\frac{1}{E-p_1'^0-\veps_n(1-i0)}
\la Pb_1 n|I(p'^0_1-p_1^0)|a_1 a_2\ra\,.
\ee
The expression (\ref{transint2}) is conveniently divided into 
 {\it irreducible} and {\it reducible} parts.
The reducible part is the one with $\veps_{Pb_2}+\veps_{n}
=E_a^{(0)}$ in first term
and with $\veps_{Pb_1}+\veps_{n} =E_a^{(0)}$ in second term.
The irredicible part is the reminder.
Using the identities
\begin{eqnarray}  \label{iden1}
 \frac{1}{p_{1}^{0}-\veps_{a_1}+i0}
       \frac{1}{E-p_{1}^{0}-\veps_{a_2}+i0}
    & =& \frac{1}{E-E_a^{(0)}}
\Bigl( \frac{1}{p_{1}^{0}-\veps_{a_1}+i0}+
       \frac{1}{E-p_{1}^{0}-\veps_{a_2}+i0}\Bigr)\,, \\
  \frac{1}{p_{1}^{\prime 0}-\veps_{P b_1}+i0}
       \frac{1}{E'-p_{1}^{\prime 0}-\veps_{P b_2}+i0} 
    & =& \frac{1} {E'-E_b^{(0)}}
\Bigl(\frac{1}{p_{1}^{\prime 0}-\veps_{P b_1}+i0}
  +    \frac{1}{E'-p_{1}^{\prime 0}-\veps_{P b_2}+i0}\Bigr)\,, 
     \label{iden2}
\end{eqnarray}
we obtain for the irreducible part
\be \label{irred1}
\tau_{\gamma_f,b;a}^{(1a,{\rm irred})}&=&
\frac{1}{2\pi i}\oint_{\Gamma_b}dE'\oint_{\Gamma_a} dE\;
g_{\gamma_f,b;a}^{(1a,{\rm irred})}(E',E)\nonumber\\
&=&\frac{1}{2\pi i}\oint_{\Gamma_b}dE'\oint_{\Gamma_a} dE\;
\frac{1}{E'-E_b^{(0)}}
\frac{1}{E-E_a^{(0)}}\nonumber\\
&& \times \Bigl\{\sum_{P}(-1)^P
\Bigl(\frac{i}{2\pi}\Bigr)^3
\int_{-\infty}^{\infty} dp^0_2 dp_2'^0
\Bigl(\frac{1}{p'^0_2-\veps_{Pb_2}+i0}
+\frac{1}{E'-p'^0_2-\veps_{Pb_1}+i0}\Bigr)\nonumber\\
&&\times
\Bigl(\frac{1}{p^0_2-\veps_{a_2}+i0}
+\frac{1}{E-p^0_2-\veps_{a_1}+i0}\Bigr)
\sum_n^{\veps_{Pb_2}+\veps_{n}
\ne E_a^{(0)}}
\la Pb_1|e \alpha_{\mu} A_f^{\mu *}|n\ra\nonumber\\
&&\times
\frac{1}{E-p_2'^0-\veps_n(1-i0)}
\la n Pb_2|I(p'^0_2-p_2^0)|a_1 a_2\ra
\nonumber\\
&&+\sum_{P}(-1)^P
\Bigl(\frac{i}{2\pi}\Bigr)^3
\int_{-\infty}^{\infty} dp^0_1 dp_1'^0
\Bigl(\frac{1}{p'^0_1-\veps_{Pb_1}+i0}+
\frac{1}{E'-p'^0_1-\veps_{Pb_2}+i0}\Bigr)
\nonumber\\
&&\times\Bigl(
\frac{1}{p^0_1-\veps_{a_1}+i0}+
\frac{1}{E-p^0_1-\veps_{a_2}+i0}\Bigr)
\sum_n^{\veps_{Pb_1}+\veps_{n}
\ne E_a^{(0)}}
\la Pb_2|e \alpha_{\mu} A_f^{\mu *}|n\ra\nonumber\\
&&\times
\frac{1}{E-p_1'^0-\veps_n(1-i0)}
\la Pb_1 n|I(p'^0_1-p_1^0)|a_1 a_2\ra\Bigr\}
\,.
\ee
The expression in the curly braces of equation (\ref{irred1})
is a regular function of $E$ or $E'$ when
$E\approx E_a^{(0)}$ and
$E'\approx E_b^{(0)}$ (see \cite{shab02} for details).
Calculating the residues and taking into account
the identity
\begin{equation}
         \frac{i}{2 \pi}\Bigl(\frac{1}{x+i0}+
                      \frac{1}{-x+i0}\Bigr)=\delta (x)\,,
    \label{e3e17}
\end{equation}
we find
\be \label{irred2}
\tau_{\gamma_f,b;a}^{(1a,{\rm irred})}&=&
-\sum_{P}(-1)^P
\Bigl\{
\sum_n^{\veps_{Pb_2}+\veps_{n}
\ne E_a^{(0)}}
\la Pb_1|e \alpha_{\mu} A_f^{\mu *}|n\ra
\frac{1}{E_a^{(0)}-\veps_{Pb_2}-\veps_n}
\la n Pb_2|I(\veps_{Pb_2}-\veps_{a_2})
|a_1 a_2\ra
\nonumber\\
&&+\sum_n^{\veps_{Pb_1}+\veps_{n}
\ne E_a^{(0)}}
\la Pb_2|e \alpha_{\mu} A_f^{\mu *}|n\ra
\frac{1}{E_a^{(0)}-\veps_{Pb_1}-\veps_n}
\la Pb_1 n|I(\veps_{Pb_1}-\veps_{a_1}
)|a_1 a_2\ra\Bigr\}
\,.
\ee
A similar calculation of the irreducible part of
the diagrams shown in Fig. 2b
yields
\be \label{irred3}
\tau_{\gamma_f,b;a}^{(1b,{\rm irred})}&=&
-\sum_{P}(-1)^P
\Bigl\{
\sum_n^{\veps_{a_2}+\veps_{n}
\ne E_b^{(0)}}
\la Pb_1 Pb_2|I(\veps_{Pb_2}-\veps_{a_2})
|n a_2\ra
\frac{1}{E_b^{(0)}-\veps_{a_2}-\veps_n}
\la n|e \alpha_{\mu} A_f^{\mu *}|a_1\ra
\nonumber\\
&&
+\sum_n^{\veps_{a_1}+\veps_{n}
\ne E_b^{(0)}}
\la Pb_1 Pb_2|I(\veps_{Pb_1}-\veps_{a_1})
| a_1 n\ra
\frac{1}{E_b^{(0)}-\veps_{a_1}-\veps_n}
\la n|e \alpha_{\mu} A_f^{\mu *}|a_2\ra\Bigr\}\,.
\ee
For the reducible part of the diagrams shown in Fig. 2a
 we have
\be \label{red1}
\tau_{\gamma_f,b;a}^{(1a, {\rm red})}&=&
\frac{1}{2\pi i}\oint_{\Gamma_b}dE'\oint_{\Gamma_a} dE \;
g_{\gamma_f,b;a}^{(1a,{\rm red})}(E',E)\nonumber\\
&&=\frac{1}{2\pi i}\oint_{\Gamma_b}dE'\oint_{\Gamma_a} dE
\frac{1}{E'-E_b^{(0)}}
\frac{1}{E-E_a^{(0)}}
\Bigl\{\sum_{P}(-1)^P
\Bigl(\frac{i}{2\pi}\Bigr)^3
\int_{-\infty}^{\infty} dp^0_2 dp_2'^0\nonumber\\
&&\times \sum_n^{\veps_{Pb_2}+\veps_{n}= E_a^{(0)}}
\Bigl[\frac{1}{E-E_a^{(0)}}
\Bigl(\frac{1}{p'^0_2-\veps_{Pb_2}+i0}
+\frac{1}{E-p'^0_2-\veps_{n}+i0}\Bigr)\nonumber\\
&&+\frac{1}{E'-p'^0_2-\veps_{Pb_1}+i0}
\frac{1}{E-p'^0_2-\veps_{n}+i0}\Bigr]
\Bigl(\frac{1}{p^0_2-\veps_{a_2}+i0}
+\frac{1}{E-p^0_2-\veps_{a_1}+i0}\Bigr)\nonumber\\
&&\times \la Pb_1|e \alpha_{\mu} A_f^{\mu *}|n\ra
\la n Pb_2|I(p'^0_2-p_2^0)|a_1 a_2\ra
\nonumber\\
&&+
\sum_{P}(-1)^P \Bigl(\frac{i}{2\pi}\Bigr)^3
\int_{-\infty}^{\infty} dp^0_1 dp_1'^0
\sum_n^{\veps_{Pb_1}+\veps_{n}= E_a^{(0)}}\nonumber\\
&&\times \Bigl[\frac{1}{E-E_a^{(0)}}
\Bigl(\frac{1}{p'^0_1-\veps_{Pb_1}+i0}
+\frac{1}{E-p'^0_1-\veps_{n}+i0}\Bigr)\nonumber\\
&&+\frac{1}{E'-p'^0_1-\veps_{Pb_2}+i0}
\frac{1}{E-p'^0_1-\veps_{n}+i0}\Bigr]
\Bigl(\frac{1}{p^0_1-\veps_{a_1}+i0}
+\frac{1}{E-p^0_1-\veps_{a_2}+i0}\Bigr)\nonumber\\
&& \times \la Pb_2|e \alpha_{\mu} A_f^{\mu *}|n\ra
\la  Pb_1 n|I(p'^0_1-p_1^0)|a_1 a_2\ra\Bigr\}\,.
\ee
Calculating the residues at $E'=E_b^{(0)}$ and
$E=E_{a}^{(0)}$ and using the identity (\ref{e3e17}),
we obtain
\be \label{tau29}
\tau_{\gamma_f,b;a}^{(1a, {\rm red})}
&=&\sum_{P}(-1)^P
\Bigl\{
\sum_n^{\veps_{Pb_2}+\veps_{n}= E_a^{(0)}}
\Bigl[\frac{i}{2\pi}
\int_{-\infty}^{\infty} dp^0_2
\frac{1}{(\veps_{a_2}-p^0_2+i0)^2}\nonumber\\
&&
\la Pb_1|e \alpha_{\mu} A_f^{\mu *}|n\ra
\la n Pb_2|I(\veps_{Pb_2}-p_2^0)|a_1 a_2\ra\Bigr] \nonumber\\
&&+
\sum_n^{\veps_{Pb_1}+\veps_{n}= E_a^{(0)}}
\Bigl[\frac{i}{2\pi}
\int_{-\infty}^{\infty} dp^0_1
\frac{1}{(\veps_{a_1}-p^0_1+i0)^2}\nonumber\\
&&
\la Pb_2|e \alpha_{\mu} A_f^{\mu *}|n\ra
\la  Pb_1 n|I(\veps_{Pb_1}-p_1^0)|a_1 a_2\ra\Bigr]
\Bigr\}\,.
\ee
We have assumed that the unperturbed states $a$ and $b$
are described by one-determinant wave functions
(\ref{ua}) and (\ref{ub}). It implies that, in equation
(\ref{tau29}), we have to consider  
$(Pb_2,n)=(a_1,a_2)$
or $(a_2,a_1)$ in first term and 
$(Pb_1,n)=(a_1,a_2)$
or $(a_2,a_1)$ in second term. 
Therefore, the reducible part contributes only in the case
 when the states $a$ and $b$ have at least one common
one-electron state. In what follows,
we assume $a_1=b_1$ and $a_2 \ne b_2$.
We obtain
\be \label{red1a}
\tau_{\gamma_f,b;a}^{(1a,{\rm red})}
&=&\frac{i}{2\pi}\int_{-\infty}^{\infty}d\omega\;
\la b_2|e\alpha_{\mu}A_f^{\mu *}|a_2\ra
 \left(\frac{\la  a_1 a_2|I(\omega)|a_1 a_2\ra}{(\omega-i0)^2}
-\frac{\la  a_2 a_1|I(\omega)|a_1 a_2\ra}{(\omega-\Delta_a-i0)^2}
\right)\,,
\ee
where $\Delta_a\equiv \veps_{a_2}-\veps_{a_1}$.
A similar calculation of the reducible part of the 
diagrams shown in Fig. 2b gives
\be \label{red1b}
\tau_{\gamma_f,b;a}^{(1b,{\rm red})}
&=&\frac{i}{2\pi}\int_{-\infty}^{\infty}d\omega\;
\la b_2|e\alpha_{\mu}A_f^{\mu *}|a_2\ra
 \left(\frac{\la  b_1 b_2|I(\omega)|b_1 b_2\ra}{(\omega-i0)^2}
-\frac{\la  b_2 b_1|I(\omega)|b_1 b_2\ra}{(\omega-\Delta_b-i0)^2}
\right)\,,
\ee
where $\Delta_b\equiv \veps_{b_2}-\veps_{b_1}$.
The reducible contribution has to be considered together
with second term in formula (\ref{transfirst1}).
Taking into account that
\be
\frac{1}{2\pi i}\oint_{\Gamma_a} dE\; g_{aa}^{(1)}(E)
&=&-\frac{i}{2\pi}\Bigl[ 2\int_{-\infty}^{\infty} dp'^0_1
\frac{1}{(p'^0-\veps_{a_1}-i0)^2} \la a_1 a_2|I(p'^0_1-\veps_{a_1})
|a_1 a_2\ra\nonumber\\
&& - \int_{-\infty}^{\infty} dp'^0_1 \frac{1}{(p'^0_1-\veps_{a_2}-i0)^2}
 \la a_2 a_1|I(p'^0_1-\veps_{a_1})
|a_1 a_2\ra\nonumber\\
&& - \int_{-\infty}^{\infty} dp^0_1 \frac{1}{(p^0_1-\veps_{a_1}-i0)^2}
 \la a_2 a_1|I(p^0_1-\veps_{a_2})
|a_1 a_2\ra
\ee
and a similar equation for the final state,
one finds
\be \label{redad}
\lefteqn{-\frac{1}{2}\oint_{\Gamma_b}dE' \oint_{\Gamma_a}dE 
\,g_{\gamma_{f},b;a}^{(0)}(E',E)\left(
\frac{1}{2\pi i}\oint_{\Gamma_a} dE\; g_{aa}^{(1)}(E)
+\frac{1}{2\pi i}\oint_{\Gamma_b} dE\; g_{bb}^{(1)}(E)\right)}
\nonumber\\
&&=\frac{1}{2}\la b_2|e\alpha_{\mu}A_f^{\mu *}|a_2\ra
\int_{-\infty}^{\infty} d\omega 
\Bigl\{
2\frac{\la a_1 a_2|I(\omega)|a_1 a_2\ra}
{(\omega-i0)^2}
+2\frac{\la b_1 b_2|I(\omega)|b_1 b_2\ra}
{(\omega-i0)^2}\nonumber\\
&& -\la a_2 a_1|I(\omega)|a_1 a_2\ra
\Bigl[\frac{1}{(\omega-\Delta_a-i0)^2}+\frac{1}{(\omega+\Delta_a-i0)^2}
\Bigr]\nonumber\\
&&-\la b_2 b_1|I(\omega)|b_1 b_2\ra
\Bigl[\frac{1}{(\omega-\Delta_b-i0)^2}+\frac{1}{(\omega+\Delta_b-i0)^2}
\Bigr]\Bigr\}\,.
\ee
Summing (\ref{red1a}), (\ref{red1b}), and (\ref{redad}), we obtain
for the total reducible contribution 
\be
\tau_{\gamma_f,b;a}^{(1,{\rm red})}&=&
-\frac{1}{2}\la b_2|e\alpha_{\mu}A_f^{\mu *}|a_2\ra
\frac{i}{2\pi}\int_{-\infty}^{\infty} d\omega 
\Bigl\{
\la a_2 a_1|I(\omega)|a_1 a_2\ra \nonumber\\
&& \times
\left[\frac{1}{(\omega+\Delta_a+i0)^2}-\frac{1}{(\omega+\Delta_a-i0)^2}
\right]\nonumber\\
&&+\la b_2 b_1|I(\omega)|b_1 b_2\ra
\left[\frac{1}{(\omega+\Delta_b+i0)^2}-\frac{1}{(\omega+\Delta_b-i0)^2}
\right]\Bigr\}\,.
\ee
Here we have employed the symmetry property of the photon propagator:
$I(\omega)=I(-\omega)$.
Using the identity
\be
\frac{1}{(\omega+i0)^2}-
\frac{1}{(\omega-i0)^2} = -\frac{2\pi}{i}\frac{d}{d\omega}
\delta(\omega)
\ee
and integrating by parts, we find
\be \label{redtot}
\tau_{\gamma_f,b;a}^{(1,{\rm red})}=
\frac{1}{2}
\la b_2 |e\alpha_{\mu} A_f^{\mu *}|a_2\ra 
[\la a_2 a_1|I'(\Delta_a) |a_1 a_2\ra
+\la b_2 b_1|I'(\Delta_b) |b_1 b_2\ra]\,,
\ee
where $I'(\Delta)\equiv \frac{dI(\omega)}{d\omega}\Bigr|_{\omega=\Delta}$
and it is implied that $a_1=b_1$.
The total expression for $\tau_{\gamma_f,b;a}^{(1)}$
(in the case $a_1=b_1$) is given by the sum of equations
(\ref{irred2}), (\ref{irred3}), and (\ref{redtot}):
\be
\tau_{\gamma_f,b;a}^{(1)}=
\tau_{\gamma_f,b;a}^{(1a,{\rm irred})}+
\tau_{\gamma_f,b;a}^{(1b,{\rm irred})}+
\tau_{\gamma_f,b;a}^{(1,{\rm red})}\,.
\ee

In addition to the interelectronic-interaction
 correction derived above,
we must take into account the contribution originating from
changing the photon energy in  the zeroth order 
transition probability (\ref{inf10ab})
 due to the interelectronic-interaction
 correction to the energies
of the bound states $a$ and $b$. It follows that the total 
interelectronic-interaction
correction to the transition probability of  first order in $1/Z$ 
is given by
\begin{eqnarray}\label{form34a}
dW^{(1)}_{\gamma_f,b;a}
=2\pi (k_f^0)^2
2{\rm Re}{\Bigl\{\tau_{\gamma_f,b;a}^{(0)*}
\tau_{\gamma_f,b;a}^{(1)}}\Bigr\}d\Omega_f+
\Bigl[
dW^{(0)}_{\gamma_f,b;a}
\Bigr|_{k_{f}^0=E_a-E_b}
-dW^{(0)}_{\gamma_f,b;a}
\Bigr|_{k_{f}^0=E_a^{(0)}-E_b^{(0)}}
\Bigr]\,,
\end{eqnarray} 
where $E_a$, $E_b$ and
$E_a^{(0)}$, $E_{b}^{(0)}$ are the energies of the bound states
$a$, $b$
with and without the interelectronic-interaction correction,
respectively.

\section{Numerical results and discussion}
\label{sec:res}
To evaluate  the
one-electron transition matrix elements, the explicit formulas 
 given in \cite{joh95} have been used. 
Infinite summations over the electron
spectrum in equations (\ref{irred2}) and (\ref{irred3})
have been performed by using the finite basis set method.
Basis functions have been constructed from B-splines by employing
the procedure proposed in \cite{bspl}. All the calculations
have been carried out for the homogeneously charged sphere model of
the nuclear charge distribution. The values for the nuclear radii 
 were taken from \cite{fbh}. 

In Tables \ref{tab:rm1}, \ref{tab:m2}, and \ref{tab:m1-exe}, 
we present our numerical results  for the decay rates
of the magnetic
transitions $2^3S_1 \rightarrow 1^1S_0$, $\ 2^3P_2 \rightarrow 1^1S_0$,
and $3^3S_1 \rightarrow 2^3S_1$, respectively. 
The values presented in the upper and lower parts of the tables have been 
obtained in the
Feynman and Coulomb gauges for the photon propagator, respectively.
The transition energies used in the calculation were taken from 
\cite{joh95,der98}.
The contribution due to the frequency dependence of the photon propagator 
($\Delta W_{\rm freq}$) 
and the  negative-continuum contribution 
($\Delta W_{e^+e^-}$)
are given in these tables as well.
It  can be seen from the tables that 
the total values of the transition probabilities in the different
gauges coincide with each other. 

As one can see from Tables \ref{tab:rm1} and \ref{tab:m2}, for the decays with $\Delta S \neq 0$,
the frequency-dependent correction is of the same and even 
larger magnitude
than the negative-continuum contribution. However, this is not the case
for the $3^3S_1 \rightarrow 2^3S_1$ transition,
 where
the correction $\Delta W_{\rm freq}$ is small compared to the
$\Delta W_{e^+e^-}$ term. The behavior of the negative-continuum
correction as a function of the nuclear charge number $Z$ agrees well with
the scaling ratio of the negative- to positive-energy contributions found
in \cite{der98} for all the transitions under consideration.  

In Tables \ref{tab:rm1comp} and \ref{tab:cmprmbpt}, we compare our results with the previous calculations
\cite{joh95,ind96,der98} that partially include the $1/Z^2$ and higher order
terms but do not account for the frequency-dependent contribution.
In Table \ref{tab:rm1comp}, the experimental data for the most
precisely measured transition $2^3S_1 \rightarrow 1^1S_0$ are 
also presented.
In the last column of this table our results are combined with the
radiative corrections that are beyond the
ones already included in the transition
energy. These corrections were recently evaluated in \cite{sap04}
for the $2s_{1/2} \rightarrow 1s_{1/2}$ transition
in hydrogenic ions for $Z\ge 50$. Since we consider high-Z two-electron ions, 
we can assume that  the one-electron (hydrogenlike) approximation is 
sufficient to evaluate the related correction in He-like ions. We have
extrapolated these data for $Z<50$ and interpolated for $Z=54$. The
uncertainties due to the extrapolation of the radiative corrections
and uncalculated
 $1/Z^2$ and higher order terms are
indicated in parentheses.
In Table \ref{tab:cmprmbpt}, the comparison with the RMBPT
calculations \cite{der98} is presented for the transitions
$\ 2^3P_2 \rightarrow 1^1S_0$ and $3^3S_1 \rightarrow 2^3S_1$.
The uncertainties due to uncalculated radiative and higher order
interelectronic-interaction corrections are also indicated.
From Tables I and IV,
it can be seen that  the  frequency-dependent contribution is 
smaller than the current experimental accuracy.

In summary, we have presented a systematic quantum electrodynamic
theory  for the
interelectronic-interaction corrections of first order in $1/Z$ to
the transition probabilities in heliumlike ions. The numerical evaluation of these
corrections to the magnetic transition probabilities has been performed and
the equivalence
of the Feynman and Coulomb gauges has been demonstrated. The results
of the calculations performed have been compared with previous RMBPT calculations
and with experiment.

\section{Acknowledgements}

Valuable discussions with I. Tupitsyn are gratefully acknowledged.
This work was supported in part by RFBR (grant no. 01-02-17248),
 by the program ``Russian Universities''
(grant no. UR.01.01.072), and by the Russian Ministry of 
Education (grant
no. E02-3.1-49). V.M.S. thanks the \'Ecole Normale Sup\'erieure for
providing support during the completion of this work. 
The work of A.V. Volotka was supported by the Russian Ministry 
of Education (grant no. A03-2.9-220).
Laboratoire Kastler Brossel is Unit\' e Mixte de Recherche du CNRS n$^\circ$ 8552.

\newpage

\begin{table}
\caption{ The decay rates of the magnetic dipole transition
$2^3S_1 \rightarrow 1^1S_0$ in units ${\rm s}^{-1}$. The 
negative-continuum contribution $\Delta W_{e^+e^-}$ and the frequency-dependent
correction $\Delta W_{\rm freq}$ are expressed in \%  with respect to the
main term $W$. $W_{\rm tot}$ is the total decay rate value. 
The values presented in the upper  part of the table were calculated
in the Feynman gauge, whereas the results presented in the lower part
were obtained using the Coulomb gauge.\label{tab:rm1}}
\begin{tabular}{cllll} \hline
$Z$&\ \ \ $W$             &$\Delta W_{e^+e^-}$ 
                                   &$\Delta W_{\rm freq}$ 
                                            &\ \ \ $W_{\rm tot}$   \\ \hline
30 &$8.9994\times 10^{ 8}$&-0.043\%&-0.029\%&$8.9929\times 10^{ 8}$\\
50 &$1.7303\times 10^{11}$&-0.08\% &-0.042\%&$1.7282\times 10^{11}$\\
70 &$5.9872\times 10^{12}$&-0.132\%&-0.045\%&$5.9766\times 10^{12}$\\
90 &$9.4551\times 10^{13}$&-0.205\%&-0.036\%&$9.4323\times 10^{13}$\\ \hline
30 &$9.0012\times 10^{ 8}$&-0.05\% &-0.042\%&$8.9929\times 10^{ 8}$\\
50 &$1.7308\times 10^{11}$&-0.09\% &-0.062\%&$1.7282\times 10^{11}$\\
70 &$5.9896\times 10^{12}$&-0.145\%&-0.073\%&$5.9766\times 10^{12}$\\
90 &$9.4596\times 10^{13}$&-0.218\%&-0.070\% &$9.4323\times 10^{13}$\\ \hline
\end{tabular}
\end{table}

\begin{table}
\caption{ The decay rates of the magnetic quadrupole transition
$2^3P_2 \rightarrow 1^1S_0$ in units ${\rm s}^{-1}$. The 
negative-continuum contribution $\Delta W_{e^+e^-}$ and the frequency-dependent
correction $\Delta W_{\rm freq}$ are expressed in \% with respect to the
main term $W$. $W_{\rm tot}$ is the total decay rate value. 
The values presented in the upper part of the table were calculated
in the Feynman gauge, whereas the results presented in the lower part
were obtained using the Coulomb gauge.\label{tab:m2}}
\begin{tabular}{cllll} \hline
$Z$&\ \ \ $W$             &$\Delta W_{e^+e^-}$ 
                                    &$\Delta W_{\rm freq}$ 
                                            &\ \ \ $W_{\rm tot}$   \\ \hline
30 &$2.1047\times 10^{10}$&-0.0001\%&0.021\%&$2.1052\times 10^{10}$\\
50 &$1.3654\times 10^{12}$&-0.001\% &0.038\%&$1.3660\times 10^{12}$\\
70 &$2.1480\times 10^{13}$&-0.005\% &0.063\%&$2.1493\times 10^{13}$\\
90 &$1.7231\times 10^{14}$&-0.017\% &0.097\%&$1.7245\times 10^{14}$\\ \hline
30 &$2.1051\times 10^{10}$&-0.0001\%&0.001\%&$2.1052\times 10^{10}$\\
50 &$1.3659\times 10^{12}$&-0.001\% &0.005\%&$1.3660\times 10^{12}$\\
70 &$2.1491\times 10^{13}$&-0.005\% &0.014\%&$2.1493\times 10^{13}$\\
90 &$1.7242\times 10^{14}$&-0.017\% &0.033\%&$1.7245\times 10^{14}$\\ \hline
\end{tabular}
\end{table}

\begin{table}
\caption{ The decay rates of the magnetic dipole transition
$3^3S_1 \rightarrow 2^3S_1$ in units ${\rm s}^{-1}$. The 
negative-continuum contribution $\Delta W_{e^+e^-}$ and the frequency-dependent
correction $\Delta W_{\rm freq}$ are expressed in \% with respect to the
main term $W$. $W_{\rm tot}$ is the total decay rate value. 
The values presented in the upper part of the table were calculated
in the Feynman gauge, whereas the results presented in the lower part
were obtained using the Coulomb gauge.\label{tab:m1-exe}}
\begin{tabular}{cllll} \hline
$Z$&\ \ \ $W$             &$\Delta W_{e^+e^-}$ 
                                  &$\Delta W_{\rm freq}$ 
                                          &\ \ \ $W_{\rm tot}$   \\ \hline
30 &$6.1245\times 10^{ 5}$&3.867\%&0.022\%&$6.3626\times 10^{ 5}$\\
50 &$1.3019\times 10^{ 8}$&2.204\%&0.034\%&$1.3311\times 10^{ 8}$\\
70 &$4.9886\times 10^{ 9}$&1.488\%&0.046\%&$5.0651\times 10^{ 9}$\\
90 &$9.0496\times 10^{10}$&1.055\%&0.059\%&$9.1503\times 10^{10}$\\ \hline
30 &$6.1273\times 10^{ 5}$&3.837\%&0.004\%&$6.3626\times 10^{ 5}$\\
50 &$1.3029\times 10^{ 8}$&2.158\%&0.006\%&$1.3311\times 10^{ 8}$\\
70 &$4.9936\times 10^{ 9}$&1.428\%&0.005\%&$5.0651\times 10^{ 9}$\\
90 &$9.0610\times 10^{10}$&0.984\%&0.002\%&$9.1503\times 10^{10}$\\ \hline
\end{tabular}
\end{table}

\begin{table}
\caption{ The decay rate (${\rm s}^{-1}$) of the transition
  $2^3S_1 \rightarrow 1^1S_0$ calculated in this work is compared
to the  previous
calculations and experiment. The experimental values and their error bars
are given in second and fourth columns, respectively.
In the last column the sum of our results and the QED
corrections obtained in \cite{sap04} are presented.
In parentheses the uncertainties of the present calculations are
indicated. Relative differences are calculated using experimental
results as a reference.\label{tab:rm1comp}}
\begin{tabular}{clllllll} \hline
$Z$&\ \ \ Exp. & Ref. & Prec. & RMBPT \cite{joh95} & MCDF \cite{ind96}
                & Present & Present+QED                   \\ \hline
23 &$5.917\times 10^{7 }$&\cite{gmm}
    &\ 4.1\%&\ \  -0.1\%&\  -0.4\%&\ 0.1\%&\ 0.0(6)\%  \\
26 &$2.083\times 10^{8 }$&\cite{gmm}
    & 12.5\%&\ \  -0.4\% &\  -0.7\%& -0.3\%& -0.4(5)\%   \\
35 &$4.462\times 10^{9 }$&\cite{dcl}
    &\ 3.2\%&\ \  -2.3\% &\  -2.5\%& -2.1\%& -2.3(4)\%   \\
36 &$5.848\times 10^{9 }$&\cite{cdl}
    &\ 1.3\%&\ \  -0.4\% &\  -0.6\%& -0.3\%& -0.5(4)\%   \\
41 &$2.200\times 10^{10}$&\cite{sbm}
    &\ 0.4\%&\ \ \ 0.8\% &\ \ 0.6\%&\ 0.9\%&\ 0.7(4)\%   \\
47 &$8.969\times 10^{10}$&\cite{bbc}
    &\ 1.8\%&\ \ \ 1.3\% &\ \ 1.2\%&\ 1.4\%&\ 1.1(2)\%   \\
54 &$3.915\times 10^{11}$&\cite{mci}
    &\ 3.0\%&\ \  -1.8\% &         & -1.6\%& -2.1(2)\%   \\ \hline
\end{tabular}
\end{table}

\begin{table}
\caption{The decay rates (${\rm s}^{-1}$)
of the transitions $2^3P_2 \rightarrow 1^1S_0$ and
$3^3S_1 \rightarrow 2^3S_1$ obtained in this work are compared
to the results obtained by RMBPT \cite{der98}. In parenthesis
the uncertainties of the present calculations are indicated. 
\label{tab:cmprmbpt}}
\begin{tabular}{cllll} \hline
   &\multicolumn{2}{c}{$2^3P_2 \rightarrow 1^1S_0$}
         &\multicolumn{2}{c}{$3^3S_1 \rightarrow 2^3S_1$}   \\
$Z$&\ \ \ This work             & RMBPT               
         &\ \ \ This work             & RMBPT               \\ \hline
30 &$2.105(4)\times 10^{10}$&$2.104\times 10^{10}$
         &$6.36(5)\times 10^{ 5}$&$6.35\times 10^{ 5}$ \\
50 &$1.366(5)\times 10^{12}$&$1.365\times 10^{12}$
         &$1.331(8)\times 10^{ 8}$&$1.33\times 10^{ 8}$ \\ 
70 &$2.149(21)\times 10^{13}$&$2.146\times 10^{13}$
         &$5.06(5)\times 10^{ 9}$&$5.06\times 10^{ 9}$ \\ 
90 &$1.724(22)\times 10^{14}$&$1.718\times 10^{14}$
         &$9.15(12)\times 10^{10}$&$9.15\times 10^{10}$ \\ \hline
                 
\end{tabular}
\end{table}

\begin{figure}
\includegraphics{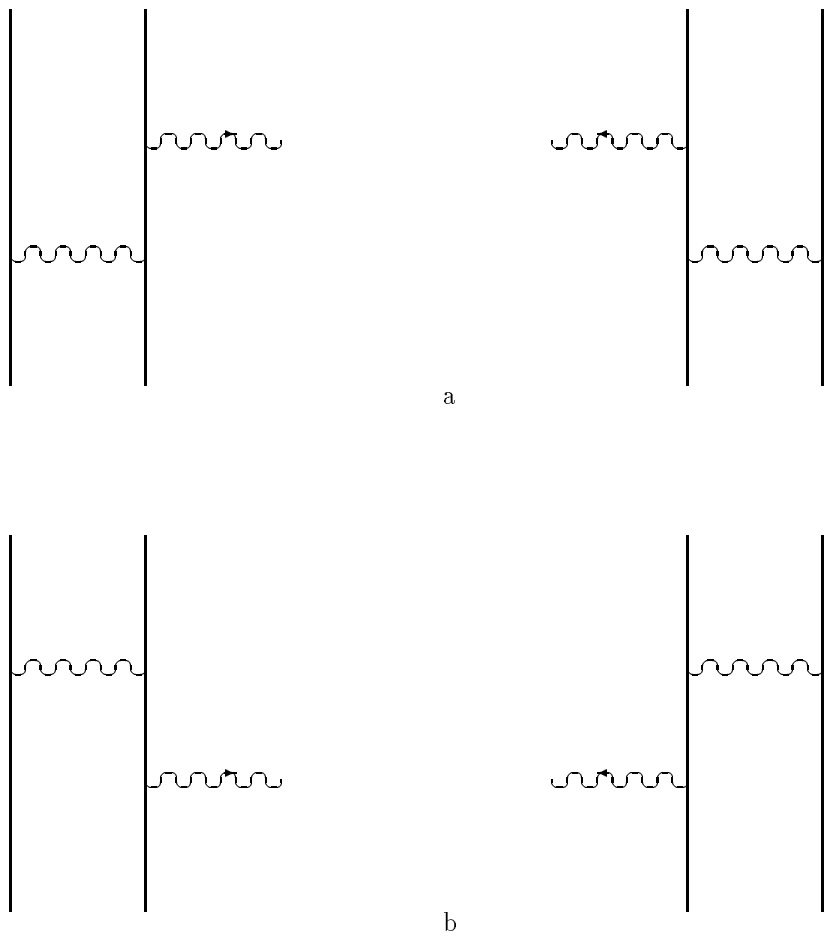}
\caption{The $1/Z$ interelectronic-interaction corrections
to the photon emission  by a heliumlike ion.\label{fig:firstz}}
\end{figure}

\begin{figure}
\includegraphics{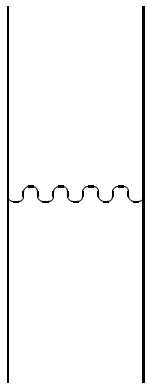}
\caption{One-photon exchange diagram.\label{onphot}}
\end{figure}

\bibliography{liter}

\begin{thebibliography}{19}
\expandafter\ifx\csname natexlab\endcsname\relax\def\natexlab#1{#1}\fi
\expandafter\ifx\csname bibnamefont\endcsname\relax
  \def\bibnamefont#1{#1}\fi
\expandafter\ifx\csname bibfnamefont\endcsname\relax
  \def\bibfnamefont#1{#1}\fi
\expandafter\ifx\csname citenamefont\endcsname\relax
  \def\citenamefont#1{#1}\fi
\expandafter\ifx\csname url\endcsname\relax
  \def\url#1{\texttt{#1}}\fi
\expandafter\ifx\csname urlprefix\endcsname\relax\def\urlprefix{URL }\fi
\providecommand{\bibinfo}[2]{#2}
\providecommand{\eprint}[2][]{\url{#2}}

\bibitem[{\citenamefont{Lindroth and Salomonson}(1990)}]{lin90}
\bibinfo{author}{\bibfnamefont{E.}~\bibnamefont{Lindroth}} \bibnamefont{and}
  \bibinfo{author}{\bibfnamefont{S.}~\bibnamefont{Salomonson}},
  \bibinfo{journal}{Phys. Rev. A} \textbf{\bibinfo{volume}{41}},
  \bibinfo{pages}{4659} (\bibinfo{year}{1990}).

\bibitem[{\citenamefont{Johnson et~al.}(1995)\citenamefont{Johnson, Plante, and
  Sapirstein}}]{joh95}
\bibinfo{author}{\bibfnamefont{W.~R.} \bibnamefont{Johnson}},
  \bibinfo{author}{\bibfnamefont{D.~R.} \bibnamefont{Plante}},
  \bibnamefont{and}
  \bibinfo{author}{\bibfnamefont{J.}~\bibnamefont{Sapirstein}},
  \bibinfo{journal}{Adv. At., Mol., Opt. Phys.} \textbf{\bibinfo{volume}{35}},
  \bibinfo{pages}{255} (\bibinfo{year}{1995}).

\bibitem[{\citenamefont{Indelicato}(1996)}]{ind96}
\bibinfo{author}{\bibfnamefont{P.}~\bibnamefont{Indelicato}},
  \bibinfo{journal}{Phys. Rev. Lett.} \textbf{\bibinfo{volume}{77}},
  \bibinfo{pages}{3323} (\bibinfo{year}{1996}).

\bibitem[{\citenamefont{Derevianko et~al.}(1998)\citenamefont{Derevianko,
  Savukov, Johnson, and Plante}}]{der98}
\bibinfo{author}{\bibfnamefont{A.}~\bibnamefont{Derevianko}},
  \bibinfo{author}{\bibfnamefont{I.~M.} \bibnamefont{Savukov}},
  \bibinfo{author}{\bibfnamefont{W.~R.} \bibnamefont{Johnson}},
  \bibnamefont{and} \bibinfo{author}{\bibfnamefont{D.~R.}
  \bibnamefont{Plante}}, \bibinfo{journal}{Phys. Rev. A}
  \textbf{\bibinfo{volume}{58}}, \bibinfo{pages}{4453} (\bibinfo{year}{1998}).

\bibitem[{\citenamefont{Lach and Pachucki}(2001)}]{lac01}
\bibinfo{author}{\bibfnamefont{G.}~\bibnamefont{Lach}} \bibnamefont{and}
  \bibinfo{author}{\bibfnamefont{K.}~\bibnamefont{Pachucki}},
  \bibinfo{journal}{Phys. Rev. A} \textbf{\bibinfo{volume}{64}},
  \bibinfo{pages}{042510} (\bibinfo{year}{2001}).

\bibitem[{\citenamefont{Pachucki}(2003)}]{pac03}
\bibinfo{author}{\bibfnamefont{K.}~\bibnamefont{Pachucki}},
  \bibinfo{journal}{Phys. Rev. A} \textbf{\bibinfo{volume}{67}},
  \bibinfo{pages}{012504} (\bibinfo{year}{2003}).

\bibitem[{\citenamefont{{V. M. Shabaev, Izv. Vuz. Fiz. \I, 43 (1990) [Sov.
  Phys. J. \I, 660 (1990)].}}()}]{shab90a}
\bibinfo{author}{\bibnamefont{{V. M. Shabaev, Izv. Vuz. Fiz. \I, 43 (1990)
  [Sov. Phys. J. \I, 660 (1990)].}}}

\bibitem[{\citenamefont{{V. M. Shabaev, Teor. Mat. Fiz. \J, 83 (1990) [Theor.
  Math. Phys. \J, 57 (1990)].}}()}]{shab90b}
\bibinfo{author}{\bibnamefont{{V. M. Shabaev, Teor. Mat. Fiz. \J, 83 (1990)
  [Theor. Math. Phys. \J, 57 (1990)].}}}

\bibitem[{\citenamefont{Shabaev}(1994)}]{shab94}
\bibinfo{author}{\bibfnamefont{V.~M.} \bibnamefont{Shabaev}},
  \bibinfo{journal}{Phys. Rev. A} \textbf{\bibinfo{volume}{50}},
  \bibinfo{pages}{4521} (\bibinfo{year}{1994}).

\bibitem[{\citenamefont{Shabaev}(2002)}]{shab02}
\bibinfo{author}{\bibfnamefont{V.~M.} \bibnamefont{Shabaev}},
  \bibinfo{journal}{Phys. Rep.} \textbf{\bibinfo{volume}{356}},
  \bibinfo{pages}{119} (\bibinfo{year}{2002}).

\bibitem[{\citenamefont{Johnson et~al.}(1988)\citenamefont{Johnson, Blundell,
  and Sapirstein}}]{bspl}
\bibinfo{author}{\bibfnamefont{W.~R.} \bibnamefont{Johnson}},
  \bibinfo{author}{\bibfnamefont{S.~A.} \bibnamefont{Blundell}},
  \bibnamefont{and}
  \bibinfo{author}{\bibfnamefont{J.}~\bibnamefont{Sapirstein}},
  \bibinfo{journal}{Phys. Rev. A} \textbf{\bibinfo{volume}{37}},
  \bibinfo{pages}{307} (\bibinfo{year}{1988}).

\bibitem[{\citenamefont{Fricke et~al.}(1995)\citenamefont{Fricke, Bernhardt,
  Heilig, Schaller, Schellenberg, Shera, and de~Jager}}]{fbh}
\bibinfo{author}{\bibfnamefont{G.}~\bibnamefont{Fricke}},
  \bibinfo{author}{\bibfnamefont{C.}~\bibnamefont{Bernhardt}},
  \bibinfo{author}{\bibfnamefont{K.}~\bibnamefont{Heilig}},
  \bibinfo{author}{\bibfnamefont{L.~A.} \bibnamefont{Schaller}},
  \bibinfo{author}{\bibfnamefont{L.}~\bibnamefont{Schellenberg}},
  \bibinfo{author}{\bibfnamefont{E.~B.} \bibnamefont{Shera}}, \bibnamefont{and}
  \bibinfo{author}{\bibfnamefont{C.~W.} \bibnamefont{de~Jager}},
  \bibinfo{journal}{At. Data and Nucl. Data Tables}
  \textbf{\bibinfo{volume}{60}}, \bibinfo{pages}{177} (\bibinfo{year}{1995}).

\bibitem[{\citenamefont{Sapirstein et~al.}(2004)\citenamefont{Sapirstein,
  Pachucki, and Cheng}}]{sap04}
\bibinfo{author}{\bibfnamefont{J.}~\bibnamefont{Sapirstein}},
  \bibinfo{author}{\bibfnamefont{K.}~\bibnamefont{Pachucki}}, \bibnamefont{and}
  \bibinfo{author}{\bibfnamefont{K.~T.} \bibnamefont{Cheng}},
  \bibinfo{journal}{Phys. Rev. A} \textbf{\bibinfo{volume}{69}},
  \bibinfo{pages}{022113} (\bibinfo{year}{2004}).

\bibitem[{\citenamefont{Gould et~al.}(1974)\citenamefont{Gould, Marrus, and
  Mohr}}]{gmm}
\bibinfo{author}{\bibfnamefont{H.}~\bibnamefont{Gould}},
  \bibinfo{author}{\bibfnamefont{R.}~\bibnamefont{Marrus}}, \bibnamefont{and}
  \bibinfo{author}{\bibfnamefont{P.~J.} \bibnamefont{Mohr}},
  \bibinfo{journal}{Phys. Rev. Lett.} \textbf{\bibinfo{volume}{33}},
  \bibinfo{pages}{676} (\bibinfo{year}{1974}).

\bibitem[{\citenamefont{Dunford et~al.}(1990)\citenamefont{Dunford, Church,
  Liu, Berry, Raphaelian, Haas, and Curtis}}]{dcl}
\bibinfo{author}{\bibfnamefont{R.~W.} \bibnamefont{Dunford}},
  \bibinfo{author}{\bibfnamefont{D.~A.} \bibnamefont{Church}},
  \bibinfo{author}{\bibfnamefont{C.~J.} \bibnamefont{Liu}},
  \bibinfo{author}{\bibfnamefont{H.~G.} \bibnamefont{Berry}},
  \bibinfo{author}{\bibfnamefont{M.~L.} \bibnamefont{Raphaelian}},
  \bibinfo{author}{\bibfnamefont{M.}~\bibnamefont{Haas}}, \bibnamefont{and}
  \bibinfo{author}{\bibfnamefont{L.~J.} \bibnamefont{Curtis}},
  \bibinfo{journal}{Phys. Rev. A} \textbf{\bibinfo{volume}{41}},
  \bibinfo{pages}{4109} (\bibinfo{year}{1990}).

\bibitem[{\citenamefont{Cheng et~al.}(1994)\citenamefont{Cheng, Dunford, Liu,
  Zabransky, Livingston, and Curtis}}]{cdl}
\bibinfo{author}{\bibfnamefont{S.}~\bibnamefont{Cheng}},
  \bibinfo{author}{\bibfnamefont{R.~W.} \bibnamefont{Dunford}},
  \bibinfo{author}{\bibfnamefont{C.~J.} \bibnamefont{Liu}},
  \bibinfo{author}{\bibfnamefont{B.~J.} \bibnamefont{Zabransky}},
  \bibinfo{author}{\bibfnamefont{A.~E.} \bibnamefont{Livingston}},
  \bibnamefont{and} \bibinfo{author}{\bibfnamefont{L.~J.}
  \bibnamefont{Curtis}}, \bibinfo{journal}{Phys. Rev. A}
  \textbf{\bibinfo{volume}{49}}, \bibinfo{pages}{2347} (\bibinfo{year}{1994}).

\bibitem[{\citenamefont{Simionovici et~al.}(1994)\citenamefont{Simionovici,
  Birkett, Marrus, Charles, Indelicato, Dietrich, and Finlayson}}]{sbm}
\bibinfo{author}{\bibfnamefont{A.}~\bibnamefont{Simionovici}},
  \bibinfo{author}{\bibfnamefont{B.~B.} \bibnamefont{Birkett}},
  \bibinfo{author}{\bibfnamefont{R.}~\bibnamefont{Marrus}},
  \bibinfo{author}{\bibfnamefont{P.}~\bibnamefont{Charles}},
  \bibinfo{author}{\bibfnamefont{P.}~\bibnamefont{Indelicato}},
  \bibinfo{author}{\bibfnamefont{D.~D.} \bibnamefont{Dietrich}},
  \bibnamefont{and}
  \bibinfo{author}{\bibfnamefont{K.}~\bibnamefont{Finlayson}},
  \bibinfo{journal}{Phys. Rev. A} \textbf{\bibinfo{volume}{49}},
  \bibinfo{pages}{3553} (\bibinfo{year}{1994}).

\bibitem[{\citenamefont{Birkett et~al.}(1993)\citenamefont{Birkett, Briand,
  Charles, Dietrich, Finlayson, Indelicato, Liesen, Marrus, and
  Simionovici}}]{bbc}
\bibinfo{author}{\bibfnamefont{B.~B.} \bibnamefont{Birkett}},
  \bibinfo{author}{\bibfnamefont{J.~P.} \bibnamefont{Briand}},
  \bibinfo{author}{\bibfnamefont{P.}~\bibnamefont{Charles}},
  \bibinfo{author}{\bibfnamefont{D.~D.} \bibnamefont{Dietrich}},
  \bibinfo{author}{\bibfnamefont{K.}~\bibnamefont{Finlayson}},
  \bibinfo{author}{\bibfnamefont{P.}~\bibnamefont{Indelicato}},
  \bibinfo{author}{\bibfnamefont{D.}~\bibnamefont{Liesen}},
  \bibinfo{author}{\bibfnamefont{R.}~\bibnamefont{Marrus}}, \bibnamefont{and}
  \bibinfo{author}{\bibfnamefont{A.}~\bibnamefont{Simionovici}},
  \bibinfo{journal}{Phys. Rev. A} \textbf{\bibinfo{volume}{47}},
  \bibinfo{pages}{R2454} (\bibinfo{year}{1993}).

\bibitem[{\citenamefont{Marrus et~al.}(1989)\citenamefont{Marrus, Charles,
  Indelicato, de~Billy, Tazi, Briand, Simionovici, Dietrich, Bosch, and
  Liesen}}]{mci}
\bibinfo{author}{\bibfnamefont{R.}~\bibnamefont{Marrus}},
  \bibinfo{author}{\bibfnamefont{P.}~\bibnamefont{Charles}},
  \bibinfo{author}{\bibfnamefont{P.}~\bibnamefont{Indelicato}},
  \bibinfo{author}{\bibfnamefont{L.}~\bibnamefont{de~Billy}},
  \bibinfo{author}{\bibfnamefont{C.}~\bibnamefont{Tazi}},
  \bibinfo{author}{\bibfnamefont{J.~P.} \bibnamefont{Briand}},
  \bibinfo{author}{\bibfnamefont{A.}~\bibnamefont{Simionovici}},
  \bibinfo{author}{\bibfnamefont{D.~D.} \bibnamefont{Dietrich}},
  \bibinfo{author}{\bibfnamefont{F.}~\bibnamefont{Bosch}}, \bibnamefont{and}
  \bibinfo{author}{\bibfnamefont{D.}~\bibnamefont{Liesen}},
  \bibinfo{journal}{Phys. Rev. A} \textbf{\bibinfo{volume}{39}},
  \bibinfo{pages}{3725} (\bibinfo{year}{1989}).

\end{thebibliography}

\end{document}